\documentclass[aps,prl,showpacs,amsfonts,amsmath,twocolumn,floatfix,superscriptaddress]{revtex4-1}
\usepackage[final]{graphicx}
\usepackage{color}

\newcommand{\vct}[1]{\mathbf{#1}}

\renewcommand\Re{\operatorname{Re}}
\renewcommand\Im{\operatorname{Im}}

\begin{document}

\title{Non-equilibrium electromagnetic fluctuations: Heat transfer and interactions}

\date{\today}

\author{Matthias Kr\"uger}
\affiliation{Massachusetts Institute of Technology, Department of
  Physics, Cambridge, Massachusetts 02139, USA}
\author{Thorsten Emig}
\affiliation{Laboratoire de Physique Th\'eorique et Mod\`eles
  Statistiques, CNRS UMR 8626, B\^at.~100, Universit\'e Paris-Sud, 91405
  Orsay cedex, France}
\author{Mehran Kardar}
\affiliation{Massachusetts Institute of Technology, Department of
 Physics, Cambridge, Massachusetts 02139, USA}

\begin{abstract}
The Casimir force between arbitrary objects {\em in equilibrium} is related to scattering from individual bodies. We extend this approach to heat transfer and Casimir forces in {\em non-equilibrium} cases where each body, and the environment, is at a different temperature. The formalism tracks the radiation from each body and its scatterings by the other objects. We discuss the radiation from a cylinder, emphasizing its polarized nature, and obtain the heat transfer between a sphere and a plate, demonstrating the validity of proximity transfer approximation at close separations and arbitrary temperatures.
\end{abstract}


\maketitle

The electromagnetic field in the space around bodies is stochastic due
to quantum and thermal fluctuations.  The basic formalism of
Fluctuational Electrodynamics (FE), was set out over 60 years ago by
Rytov~\cite{Rytov3}, and has been applied extensively since to diverse
problems in radiative heat transfer~\cite{Polder71,Modest} and Casimir
forces~\cite{Lifshitz56}.  FE starts with casting the current
fluctuations in each body in terms of its dielectric properties, and
proceeds to compute the resulting electromagnetic field.  The improved
precision of measurements of force and heat transfer at sub-micron
scale have
provided renewed incentive to examine FE for objects at different
temperatures \cite{Bordag}. In particular, when the size
 or separation of 
the objects is comparable to, or smaller than, the thermal wavelength
(around 8 micron at room temperature), heat radiation and transfer
will differ from the predictions of 
the Stefan-Boltzmann law: The considerably
larger near-field heat transfer, due to tunneling of evanescent waves,
has been verified experimentally~\cite{Sheng09, Rousseau09}.
Theoretical computations of  heat transfer were only recently
extended from two parallel plates~\cite{Polder71} or
dipoles~\cite{Volokitin01} to two spheres~\cite{Narayanaswamy08}.  The
radiation of single spheres and plates has been studied by many
authors \cite{Kattawar70,Bohren}. For a cylinder, the emissivity restricted to waves traveling perpendicular to its axis has been addressed \cite{Bimonte09b}. There are also recent computations
of the non-equilibrium Casimir force between objects at different
temperatures, for parallel plates~\cite{Antezza08}, modulated
plates~\cite{Bimonte09}, as well as a plate and an atom
\cite{Antezza05}.  The limitation of these results to simple shapes
and arrangements points out the need for approaches capable of
handling more complex situations.

Here, we derive a formalism for computing heat transfer and Casimir forces for arbitrary objects (compact or not) maintained at different temperatures. 
Generalizing previous work on Casimir forces in equilibrium, our approach enables
systematic description of FE of a collection of objects in terms of their individual scattering properties.
For the non-equilibrium Casimir force, we can investigate interactions between
compact objects where, unlike previous studies~\cite{Antezza05,Antezza08,Bimonte09}, 
the effect of a third temperature (of the environment) has to be taken into account. 
In terms of new applications, we derive the heat radiation of a cylinder which is of interest for heated wires or carbon nanotubes \cite{Fan09}. We also study the heat transfer between a sphere and a plate, the only geometry for which near field heat transfer has been measured~\cite{Sheng09, Rousseau09}.

Consider an arrangement of $N$ objects labelled as $\alpha=1\dots N$,
in vacuum at constant temperatures $\{T_\alpha\}$, and embedded in an
environment at temperature $T_{env}$.  In this non-equilibrium stationary
state, each object is assumed to be at local
equilibrium with current fluctuations obeying the
fluctuation-dissipation theorem (FDT).  In the following, we derive
the autocorrelation function $C$ of the electric field $\vct{E}$ at
frequency $\omega$ at points $\vct{r}$ and $\vct{r}'$ outside the objects, from which the Poynting
vector for heat transfer and the Maxwell stress tensor for Casimir forces
can then be extracted.  In equilibrium, with
$T_\alpha=T_{env}=T$,
$C$ is related to the imaginary part of the dyadic Green's function
$G_{ij}$ by~\cite{Rytov3,Eckhardt83},
\begin{align}
\notag C_{ij}^{eq}(T)&\equiv\left\langle E_i(\omega;\vct{r})E_j^*(\omega;\vct{r}') \right\rangle^{eq}\\&=\left[a_T(\omega)+a_0(\omega)\right] \frac{c^2}{\omega^2}\Im G_{ij}(\omega;\vct{r},\vct{r}'),\label{eq:1}
\end{align}
where $a_T(\omega)\equiv \frac{\omega^4 \hbar
  (4\pi)^2}{c^4}(\exp[\hbar \omega/k_BT]-1)^{-1}$ is proportional to the occupation
number of all oscillators of frequency $\omega$, $c$ is the speed of light, and
$\hbar$ is Planck's constant.  Zero point fluctuations which
contribute $a_0(\omega)\equiv \frac{\omega^4 \hbar (4\pi)^2}{2c^4}$
play no role in our discussion.  We shall henceforth employ the
operator notation $\mathbb{G}\equiv G_{ij}(\omega ;\vct{r},\vct{r}')$.  Since
$\Im \mathbb{G}=-\mathbb{G} \Im \mathbb{G}^{-1}\mathbb{G}^*$, and
using the identity \cite{Eckhardt83} $\sum_\alpha\Im \varepsilon_\alpha \mathbb{I}=-
\frac{c^2}{\omega^2}\Im (\mathbb{G}^{-1}-\mathbb{G}_{0}^{-1})$,
where $\varepsilon_\alpha$ is the complex dielectric function of
object $\alpha$ and $\mathbb{G}_{0}$ is the Green's function of free
space, we obtain
\begin{align}
C^{eq}(T)&=C_0+\sum_\alpha C_\alpha^{sc}(T)-a_T(\omega) \frac{c^2}{\omega^2}\mathbb{G} \Im \mathbb{G}_0^{-1} \mathbb{G}^*\label{eq:4}\notag,\\
C_\alpha^{sc}(T)&=a_T(\omega)\mathbb{G} \Im \varepsilon_\alpha \mathbb{G}^*,
\end{align}
where $C_0=a_0(\omega)\frac{c^2}{\omega^2}\Im \mathbb{G}$ is the zero
point term.  The finite temperature contribution is thus a
sum of $N+1$ terms: Each $C_\alpha^{sc}(T)$ contains an implicit
integral over sources within $\alpha$ and is identified with the field
sourced by this object~\cite{Rytov3}; the scattering of this radiation
by all other objects is accounted for by multiplying $\Im
\varepsilon_\alpha$ on both sides with the full Green's function.  The
last term in Eq.~\eqref{eq:4},
$C^{env}(T)=-a_{T}(\omega)\frac{c^2}{\omega^2}\mathbb{G} \Im \mathbb{G}_0^{-1}
\mathbb{G}^*$, is hence identified with the contribution sourced
  by 
the environment. 

A key assumption of FE is that in a non-equilibrium situation, the
thermal current fluctuations inside each object are described by the
FDT at the corresponding local temperature, and are independent of the
impinging radiation from the other objects.  Having identified the
different sources in Eq.~\eqref{eq:4}, we can change their
temperatures to arrive at the desired non-equilibrium generalization
\begin{align}
C^{neq}(T_{env},\{T_\alpha\})=C_0+\sum_\alpha C_\alpha^{sc}(T_\alpha)+C^{env}(T_{env})\notag\\
=C^{eq}(T_{env})+ \sum_\alpha \left[ C_\alpha^{sc}(T_\alpha)-
  C_\alpha^{sc}(T_{env}) \right]
\label{eq:6}. 
\end{align}
The second form is obtained by considering
the {\it difference} of $C^{neq}(T_{env},\{T_\alpha\})$ from
$C^{eq}(T_{env})$ due to the {\it deviations} of the object
temperatures $T_\alpha$ from $T_{env}$.  This form is useful because
the equilibrium correlation can be regarded as known, and the number
of sources is reduced from $N+1$ to $N$.  Applying the formalism,
e.g., to derive Casimir forces, the first term on the r.h.s. of
Eq.~\eqref{eq:6} yields the equilibrium force at temperature
$T_{env}$. 

The next step is to compute the radiation field of object $\alpha$
when {\em isolated}, i.e., before this field is scattered by the other
objects, and with $T_{env}=0$. This is given by
$C_\alpha(T_{\alpha}) \equiv a_{T_\alpha}(\omega)\mathbb{G}_\alpha \Im \varepsilon_\alpha \mathbb{G}_\alpha^*\,$
where $\mathbb{G}_\alpha$ is the Green's
function of object $\alpha$ in isolation, and thus involves an
implicit integration over the interior of object $\alpha$.  To employ
multiple scattering techniques~\cite{Rahi09}, it is considerably more
convenient to express $C_\alpha(T_{\alpha})$ in terms of the
T-operator or scattering amplitude $\mathbb{T}_\alpha$ of the object.  In equilibrium, the electric field
correlator for the isolated object $C_\alpha^{eq}(T_{\alpha})=
a_{T_\alpha}(\omega)\frac{c^2}{\omega^2}\Im\mathbb{G}_\alpha$,
contains radiation sourced (i) by the environment and (ii) by the
object itself.  The latter can be obtained by subtracting the
contribution from the environment, which can be regarded as an
additional material with $\varepsilon_{env}\to 1$, occupying the space
complimentary to $\alpha$~\cite{Eckhardt83}.  Towards this
calculation, we introduce a Green's function
$\tilde{\mathbb{G}}_\alpha$ with $\varepsilon_\alpha$ inside object
$\alpha$ and $\varepsilon_{env}$ outside, 
\begin{eqnarray}
C_\alpha(T_{\alpha})&=&C_\alpha^{eq}(T_{\alpha})-C_\alpha^{env}(T_{\alpha}),\label{eq:ob}
\\\notag
C_\alpha^{env}(T_{\alpha})&=&a_{T_\alpha}(\omega)\lim_{{\varepsilon_{env}}\to1}  \tilde{\mathbb{G}}_{\alpha} \Im {\varepsilon_{env}} ~\tilde{\mathbb{G}}^*_{\alpha}.
\end{eqnarray}
Note that all sources for $C_\alpha^{env}(T_{\alpha})$ are outside
object $\alpha$, and none of the Green's functions appearing in
Eq.~(\ref{eq:ob}) contain points
inside the object, which can thus be written in terms of
$\mathbb{T}_\alpha$ as
$\mathbb{G}_\alpha=\mathbb{G}_0-\mathbb{G}_0\mathbb{T}_\alpha\mathbb{G}_0$~\cite{Rahi09} ($\tilde{\mathbb{G}}_\alpha$ is a simple modification of
$\mathbb{G}_\alpha$ as a finite $\varepsilon_{env}-1$ only changes the
external speed of light).
For computing the energy radiated by object $\alpha$, one does not
have to find $C_\alpha^{eq}(T_{\alpha})$: As a consequence of detailed
balance, it does not contribute to the Poynting vector.

Finally, to compute $C_\alpha^{sc}(T_{\alpha})$ in
Eq.~\eqref{eq:6}, we need to account for scattering of the radiation
emerging from $\alpha$, by all other objects collectively designated
by $\beta$.  
Denoting their total T-operator by $\mathbb{T}_\beta$, by use of
the Lippmann-Schwinger equation~\cite{Rahi09}, we arrive at the final
form
\begin{eqnarray}
C_\alpha^{sc}(T_\alpha)&=&\mathbb{O}_{\alpha,\beta}\,C_\alpha(T_\alpha) \,\mathbb{O}_{\alpha,\beta}^\dagger\,,\quad{\rm with}\,\label{eq:ms}\\
\mathbb{O}_{\alpha,\beta}&=& (1-\mathbb{G}_0\mathbb{T}_\beta)\frac{1}{1-\mathbb{G}_0\mathbb{T}_\alpha\mathbb{G}_0\mathbb{T}_\beta}\, .\notag
\end{eqnarray}
Expanding the resolvent leads to an alternating 
application of $\mathbb{T}_\beta$ and $\mathbb{T}_\alpha$, 
corresponding to a sequence of scatterings between the objects. 
Equations~\eqref{eq:ms} and \eqref{eq:6} constitute our non-equilibrium formalism. 

The correlator $C^{neq}$ enables computing the Poynting vector and Maxwell stress tensor,
respectively given by
\begin{align}
\vct{S}(\vct{r})&=\frac{c}{4\pi}\int\frac{d\omega}{(2\pi)^2}\left\langle\vct{E}(\omega,\vct{r})\times\vct{B}^*(\omega,\vct{r})\right\rangle\,\notag,\\
T_{ij}(\vct{r})&=\int \frac{d\omega}{16\pi^3}\left\langle E_i E^*_j+B_i B^*_j-\frac{1}{2}\left(|E|^2+|B|^2\right)\delta_{ij}\right\rangle,\notag
\end{align}
where the arguments $\omega$ and $\vct{r}$ are omitted in the lower
line.  The heat $H$ absorbed  per unit time by object $\alpha$, and the force $F_i$ acting on this object in
direction $i$, are then obtained by integrations of $\vct{S}$ and
$T_{ij}$ over a surface $\sigma_\alpha$ enclosing {\it only} this
object, as
\begin{equation}
H_\alpha=-\Re\oint_{\sigma_\alpha} \vct{S}\cdot \vct{n}_{\alpha} \, d\sigma, \,\,\, F_{i,\alpha}=\Re\oint_{\sigma_\alpha}  T_{ij}n_{\alpha,j} \,d\sigma , \label{eq:fin}
\end{equation}
where $\vct{n}_{\alpha}$ is the outward normal to the surface $\sigma_\alpha$. 

\begin{figure}
\includegraphics[width=0.99\linewidth]{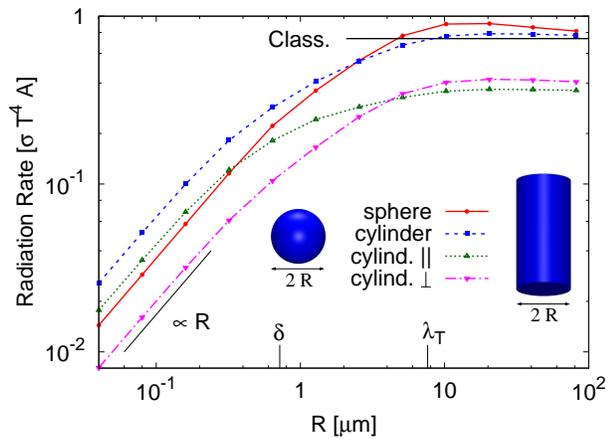}
\caption{\label{fig:1}
Heat radiation of a cylinder and a sphere of SiO$_2$, as function of  $R$,  
normalized by the Stefan Boltzmann result, at  $T=300 K$. 
The horizontal line shows the radiation of a SiO$_2$ plate.
$\lambda_T$ and the smallest skin depth $\delta$ in the relevant
frequency range are marked on the $R$-axis. 
For the cylinder,  the contributions of the different
polarizations are shown.} 
\end{figure}

As a first application we compare heat radiations from a single
object, a plate, sphere or cylinder; the only shapes amenable to
analytic treatment. As formulae for heat
radiation of a sphere or a plate are available in the
literature~\cite{Kattawar70, Rytov3}, we focus on the cylinder where
the corresponding result is only discussed implicitly~\cite{Rytovc}.
  For an infinitely long cylinder, $\mathbb{T}$ is
represented in cylindrical wave functions~\cite{Rahi09}, indexed by
$(n,k_\|,P)$ where $n$ is the multipole order,
  $k_\|$ the wave vector component along the cylinder, and $P=E$ or $P=M$
  the polarization.  The matrix element $T^{P'P}_{n,k_\|}$ describes the
relative amplitude of the scattered wave of mode $(n,k_\|,P')$ emerging from
an incoming wave $(n,k_\|,P)$.  We then find for the radiated heat
 of the single cylinder per length $L$,
\begin{eqnarray}
\notag\frac{|H_c|}{L}&&= -\int_0^\infty\frac{d\omega}{(2\pi)^2} a_{T}(\omega)\frac{c^4}{4\pi^2\omega^3}\sum_{P=E,M}\sum_{n=-\infty}^{\infty} \\
&&\int\limits_{-\omega/c}^{\omega/c}dk_{||}
\left(\Re[T_{n,k_{||}}^{PP}]+|T_{n,k_{||}}^{PP}|^2+
|T_{n,k_{||}}^{P\bar P}|^2\right)\label{eq:radcyl},
\end{eqnarray} 
where $\bar P=M$, if $P=E$ and vice versa.  In Fig.~\ref{fig:1}, we
compare the heat radiation of a plate (semi-infinite body), a sphere
and a cylinder (both of radius $R$), all evaluated with optical data
of silicon-dioxide (SiO$_2$), as used in experiments~\cite{Sheng09}.
The radiation is normalized to the
  Stefan-Boltzmann law $H=\sigma T^4 A$, with 
$\sigma=\pi^2 k_B^4 / (60 \hbar^3 c^2)$ and
surface area $A$ of the object.

For thin cylinders and small spheres $H$ is proportional to the
volume, while in the opposite limit, it is proportional to the surface
area, reflecting a finite skin depth (absorption length)
$\delta(\omega)=c/(\Im\sqrt{\varepsilon}\omega)$:
Thermal fluctuations at frequency $\omega$ within the object emit
radiation which may be re-absorbed on its way out.  If $\delta\ll R$,
only thermal fluctuations near the surface lead to  emerging
radiation, while for $\delta\gg R$, the entire volume contributes to
$H$.  An interesting feature of Fig.~\ref{fig:1} is the intermediate
range, where the sphere and cylinder emit more strongly than a plate
of equal area, related to Mie resonances for the sphere \cite{Bohren}.  For $R\to\infty$, i.e., when the wavelengths involved
(roughly peaked around the thermal wavelength $\lambda_T=\hbar
c/k_BT\approx 7.6 \mu$m), as well as skin depths, are much
smaller than the smallest dimension of the object, the classical
(plate) result is approached. The asymptotic value is in these units
denoted as emissivity $e(T)<1$.
Interestingly, the radiation from a cylinder is polarized, its
parallel and perpendicular polarizations obtained from the $P=E$ and
$P=M$ terms in Eq.~\eqref{eq:radcyl}, respectively. 
The predominant radiation of a thin cylinder
is parallel and changes to perpendicular for
$R\approx\lambda_T$. Both polarizations
  become equal asymptotically as $R\to\infty$.
Polarization effects have indeed been observed experimentally for wires \cite{Ohman61,Bimonte09b} and carbon-nano-tubes~\cite{li03}, for which other
descriptions have been offered~\cite{Aliev2008}. 

Now we consider multiple objects.
To compute $C^{neq}$ involving spheres or cylinders, we also
need to convert among bases appropriate to the different objects.  For
example, for the experimentally most relevant configuration of a
sphere and a plate~\cite{Sheng09}, the radiation from the plate, given
in a plane wave basis, must be transformed to the spherical
basis~\cite{Rahi09}, reflected by the sphere, transformed back and so
on.  For simplicity, we focus on a plate held
at a finite temperature $T_{p}\not= 0$, while the sphere and
environment are at zero temperature ( $T_{s}=T_{env}=0$). This
suffices to describe also  situations with
$T_{s}\not= 0$ as the transfer vanishes for $T_p=T_s$, and hence for $T_s\not= 0$, we subtract our result evaluated at $T_s$.
We express the correlations in
Eq.~\eqref{eq:6} in a plane waves basis, and $H_s$, the energy absorbed
by the sphere, is obtained by integrating $\vct{S}$ in
Eq.~\eqref{eq:fin} over two infinite parallel
  planes enclosing the sphere and separating it from the plate.

\begin{figure}
\includegraphics[width=0.99\linewidth]{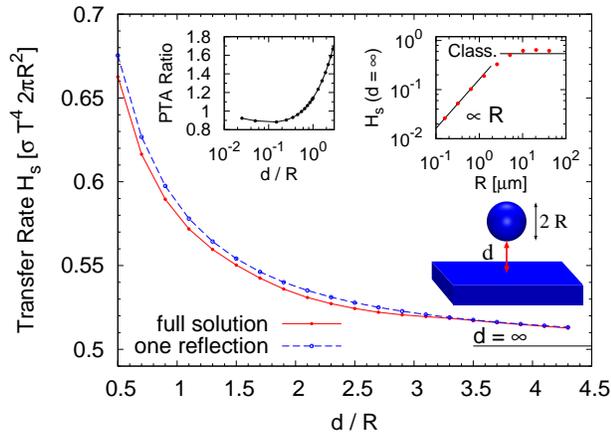}
\caption{\label{fig:2} Heat transfer rate (in units of
  Stefan-Boltzmann's law) from a
  room temperature plate to a sphere at $T=0$ of radius $R=5\mu$m (both
  SiO$_2$), as a function of separation.  The horizontal bar shows the
  limit of $d\to\infty$.  The left inset shows the approach to PTA for
  the divergent terms in a one reflection approximation.  The right
  inset shows the result at large separation as function of $R$.}
\end{figure}

Figure~\ref{fig:2} shows the results for the heat transfer from a
SiO$_2$ plate at room temperature to a SiO$_2$ sphere of $R=5\mu$m at
zero temperature, with surface-to-surface separation $d$,
normalized by the Stefan-Boltzmann law
$H_s=\sigma T^4 2\pi R^2$ (only half of the sphere is exposed to the
plate).  For large $d$, $H_s$ is roughly 0.5 in these units, whereas
for $d\to0$, $H_s$ diverges due to the increased tunneling of
evanescent waves, eventually exceeding the Stefan-Boltzmann value. The figure shows the numerical
solution of Eq.~\eqref{eq:6} together with a one reflection
approximation, where we set
$\mathbb{O}_{\alpha,\beta}=(1-\mathbb{G}_0\mathbb{T}_\beta)$ 
 in Eq.~\eqref{eq:ms}, neglecting higher order reflections between
 sphere and  plate. We see that the two curves approach each
other for large $d$, as most rays are scattered outward and will not
hit the sphere a second time. The reflection expansion is hence
helpful for getting analytical results for $d\gg R$.  Our numerical
solution involves an expansion in spherical multipoles: For
$R/\lambda_T$ large or $d/R$ small, 
increasingly more multipoles are
needed.  In practice, we restrict to a maximal multipole order of
$l_{max}=20$, for accurate results up to $d\geq R/2$.  Since closer
separations are also interesting and relevant experimentally, but
difficult numerically, we demonstrate in the inset of Fig.~\ref{fig:2}
the approach to a proximity transfer approximation (PTA), equivalent
to the proximity force approximation (PFA) used in Casimir physics,
\begin{equation}
\lim_{d/R\to0} H_s(d)=2\pi R\int_d^{d+R} S^{pp}(s) \, ds\,, \label{eq:PTA}
\end{equation}
where $S^{pp}(d)$ is the Poynting vector for parallel plates at separation $d$.
We identify the divergent terms as $d\to0$ for both the sphere-plate
and plate-plate configurations (the $E$ modes originating from
evanescent waves), and evaluate their ratio in the one reflection
approximation (allowing us to use $l_{max}=200$).  As demonstrated in
Fig.~\ref{fig:2}, this ratio approaches unity
for $d\to0$, suggesting $\mathbb{T}$ of the sphere approaches
  $\mathbb{T}$ of the plate in the PTA-sense. 
 From this, we anticipate that
multiple applications of these matrices (leading to the full solution)
will also approach the ratio unity (independent of the accuracy of the
one reflection approximation as $d\to0$). 
   We investigated different
$R$ and confirmed that PTA in Eq.~\eqref{eq:PTA} is valid in general,
with $H_s(d)\propto d^{-1}$ as $d\to0$.  While a similar point is
discussed in Ref.~\cite{Narayanaswamy08} for the case of two spheres
and used in experimental studies \cite{Sheng09,Rousseau09}, to our
knowledge the validity of PTA was not quantitatively demonstrated
previously. It is not obvious as it implies that the ratios
$R/\lambda_T$ and $R/\delta$ are irrelevant as $d\to0$.

For $d\to\infty$, $H_s$ approaches a constant, which can be obtained
by considering the $d$ independent part of the plate radiation, and
using the one reflection term.  The result, as shown in the right
inset of Fig.~\ref{fig:2}, is quite similar to the behavior of a single sphere in
Fig.~\ref{fig:1}: For small $R$, $H_s$ is proportional to the volume
of the sphere, for similar reasons as discussed before. In this limit,
$H_s$ is given by (with magnetic permeability of the sphere $\mu_s$
and Fresnel reflection coefficients $r^E$ and $r^M$ of the plate for angle $\theta$),
\begin{align}
\notag\lim_{d\gg \lambda_T\gg R}H_s&=\frac{cR^3}{16\pi^3}\int_0^\infty \!d\omega a_T(\omega) \Im\left(\frac{\mu_s-1}{\mu_s+2}+\frac{\varepsilon_s-1}{\varepsilon_s+2}\right)\\&\int_{0}^{\pi/2}d\theta \sin\theta\sum_{P=E,M}(1-|r^P(\theta,\omega)|^2).
\end{align}
For $R\gg \lambda_T$, we may expect the result to approach a classical
limit, given by $\sigma T^4 e^2 2\pi R^2$, with $e$ from Fig.~1.
While the data points come close to this value, one does not expect 
exact approach~\cite{Modest}, in contrast to Fig.~\ref{fig:1}, because
the Fresnel coefficients depend on the angle of incidence.  If we
additionally let $(\varepsilon_p, \varepsilon_s)\to1$,
$H_s$ will approach the classical limit since the
  Stefan-Boltzmann law applies to all convex black bodies.

While we highlighted applications to simple shapes, the formalism
presented here is more general, and combined with a numerical scheme
for the computation of scattering matrices~\cite{Reid} can deal with
collections of objects at different temperatures.  Indeed, such a
formalism is needed to properly deal with near field effects in device
and fabrication at the micron scale.  The formalism 
 yields also Casimir forces between objects at different temperatures-
examples of which we leave for future work.  We note,
however, that in the final stages of this project we became aware of
two independent, partly related, studies of non-equilibrium effects~\cite{Messina,Otey11}.

\begin{acknowledgments}
This research was supported by the DFG grant No. KR 3844/1-1, NSF Grant No.  DMR-08-03315 and DARPA contract No. S-000354.
We thank G. Bimonte, R.L. Jaffe, M.F. Maghrebi and G. Chen for discussions, and P. Sambegoro for providing optical data.
\end{acknowledgments}

%

\end{document}